\documentclass[%
 aip,
rsi,%
 amsmath,amssymb,
 reprint,%
]{revtex4-1}

\usepackage{graphicx}% Include figure files
\usepackage{dcolumn}% Align table columns on decimal point
\usepackage{bm}% bold math
\usepackage{color}
\begin{document}

\preprint{AIP/123-QED}

\title{Quantifying Dynamical Heterogeneity Length Scales of Interface Water across Model Membrane Phase Transition}%:\\with Forced Linebreak\footnote{Error!}}% Force line breaks with \\
% \thanks{Footnote to title of article.}

% \author{A. Author}
%  \altaffiliation[Also at ]{Physics Department, XYZ University.}%Lines break automatically or can be forced with \\
\author{Sheeba Malik}
\affiliation{Department of Chemistry, IIT Jodhpur, Jodhpur, Rajasthan, India}
\author{Smarajit Karmakar}
\affiliation{Centre for Interdisciplinary Sciences, Tata Institute of Fundamental Research, Hyderabad, India}
%\email{smarajit@tifrh.res.in}
\author{Ananya Debnath}
\affiliation{Department of Chemistry, IIT Jodhpur, Jodhpur, Rajasthan, India}
\email{ananya@iitj.ac.in}
%\author{Arpita Srivastava and Ananya Debnath}%
% \email{ananya@iitj.ac.in}
%\affiliation{ Department of Chemistry, Indian Institute of Technology Jodhpur, Rajasthan, India.}


\begin{abstract}
All-atom molecular dynamics simulations of 1,2-dimyristoyl-sn-glycerol-3-phosphocholine lipid membranes reveal a membrane phase transition dictated drastic growth in the interface water (IW) heterogeneity length scales. It acts as an alternate probe to capture the ripple size of the membrane and follows an activated dynamical scaling with the relaxation time scale solely within the gel phase. The results quantify the mostly unknown correlations between the spatio-temporal scales of the IW and membranes at various phases under physiological and supercooled conditions
\end{abstract}
%%%MAIN TEXT%%%%
\maketitle
\begin{figure}
\centering
 \includegraphics[width=\linewidth]{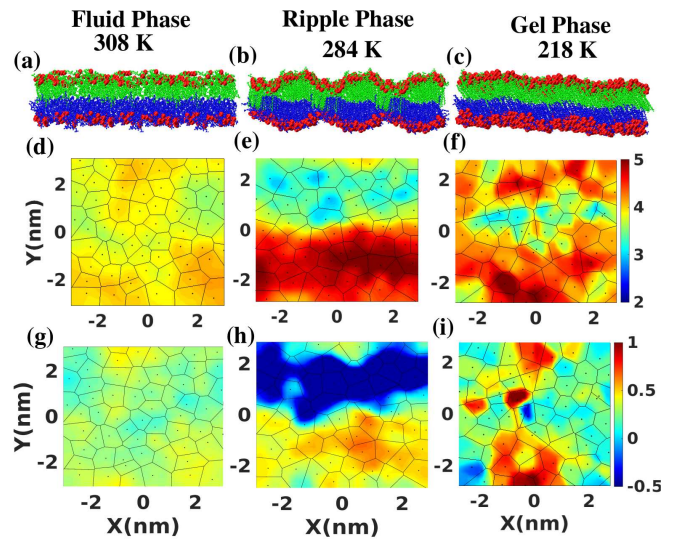}
 \caption{Snapshots of bilayers at three phases (a-c), thickness (d-f) and interdigitation (g-i) superimposed on voronoi area per lipid head respectively. Color code with representation: DMPC: CPK representation in blue and green for two leaflets; IW water: VDW (van der Waals) representation in red.}
 \label{snapshot}
\end{figure}

%%%%%%%%%%%%%%%%%%%%%%%%%%%%%%%%%%%%%%%%%%%%%%%%%%%%%%%%%%%%%%%%%%%%%
Lipid bilayers form a continuous semipermeable barrier surrounding cells and compartmentalize the intracellular space. They are essential for controlling the dynamics, arrangement, and function of membrane proteins \cite{smith2012lipid} at the fluid (or liquid-crystalline, L$_\alpha$) phase at physiological temperature. The bilayers at the fluid phase comprise of heterogeneous functional microdomains that aid in activities including cell signalling, cell adhesion, and membrane trafficking. Thus, the lateral heterogeneity is accepted as a criterion for the function of a complex biological membrane, specifically in the context of a raft \cite{jacobsonNature:2007} relevant for signalling cascades \cite{ToomreNature:2000}. Quantifying the heterogeneity in a bio-membrane using the current biochemical or biophysical tools is an incredibly challenging task due to the presence of unusual liquid-like domains with microseconds to seconds time scale and nanometers to micrometers length scales having both short and long-range orders. 
Molecular level characterization of such dynamical domains through experiments is still fragmentary due to the lack of the relevant atomistic trajectories and the unknown spatio-temporal correlations. \\
At physiological temperature interface water molecules (IW) near a single component fluid membrane exhibit signatures of dynamical heterogeneity \cite{DebnathSM:2019} and can probe regional lipid dynamics which is difficult to access otherwise \cite{DebnathPCCP:2020}. This is the stepping stone of the current study which aims to measure the spatio-temporal scales of different phases of functionally relevant bilayers through the IW dynamics. Using sum-frequency generation spectroscopy heterogeneity of IW at the water-charged lipid interface has been identified \cite{Bonn_JPCL_2019}. The fluid or disordered membrane undergoes a phase transition to a ripple (P$_\beta$) phase which quickly changes to a gel (L$_\beta$) phase \cite{debnathSM:2014} upon cooling. The membrane remains in the gel phase upon supercooling. Similar disorder to order transition of the membrane occurs from dehydration induced drastic slow-down in structural relaxation originated from the dynamical heterogeneity of the interface water molecules (IW) \cite{DebnathJCP:2021}.
Understanding cellular membranes at supercooled temperatures is extremely crucial for cryo-preservation methods, which have a
wide range of applications in food science, pharmacology,
clinical medicine, and the preservation of human embryos
for in vitro fertilization \cite{LoisNature:2000}.
Despite such innumerable applications,
investigations on the lipid dynamics at the gel phase are rarely found \cite{starrJPCB:2016}. Using QENS, higher contribution of rotational water dynamics is found near multilamellar DMPC bilayers compared to the translational one at supercooled temperature \cite{swenson2008solvent}. Although it is widely accepted that IW aids in different activities of cellular membranes, very few works attempt to establish the correlations of hydration dynamics to membrane functions \cite{ChattopadhyayJPCL:2021}. Using coarse-grained (CG) and all-atom (AA) molecular simulations and different experimental techniques, the phase behavior of model membranes have been investigated extensively \cite{Debnath_JPCB_20,KlaudaBBA:2018,OldfieldBiochemistry:1979, SturtevantPNAS:1976, HansmaScience:1988, akabori:2015}. However, none of the work mentioned above focuses on the underlying role of dynamical heterogeneity across phase transitions.
 The thermodynamics and dynamics of the IW exhibit sharp cross-overs across the bilayer melting transition which demonstrates their correlations \cite{debnathPRL:2013}. Thus, the effect of {\it membrane} phase transitions on dynamical heterogeneity of the IW is fundamentally
important and long overdue.
Although many theoretical models \cite{Sarika_PNAS_08} and numerical methods are developed to study glass transition phenomena, their applicabilities on the membrane phase transitions and water dynamics are not understood. The current research finds out the relevance of different theories \cite{Thirumalai_PRL_1987} and recent numerical approaches \cite{karmakarPNAS:2009} of glass dynamics on model membranes across phase transitions. The objective is many fold: a) to gain a fundamental understanding of the dynamical heterogeneity of the IW and the lipid membranes near phase transitions, b) to establish relations of the spatio-temporal scales of the IW to that of the membranes, c) to build up an alternate method through IW dynamics to probe membrane domains under physiological and stressed conditions. Furthermore, this strategy will help in comprehending the domain length scale of a raft, which is rarely found in theory and extremely difficult to gain from an experiment.\\
Therefore, $11.55~\mu$s long all-atom molecular dynamics (MD) simulations of dimyristoylphosphatidylcholine (DMPC) bilayers are performed at various temperatures and two different system sizes ( $6\times6$ nm$^2$ and $18\times18$ nm$^2$) to cover the three primary phases.
Simulation details are mentioned in section I of the Supporting Information (SI) and the runlengths of all bilayers are mentioned in table S1 of the SI. CHARMM36  force fields are used for the 128 DMPC lipids solvated in $5743$ TIP4P/2005 \cite{JLF:2005}($6\times6$ nm$^2$ box size) water molecules. This particular pairing of the lipid force-fields and the water model accurately reproduces the fluid to ripple to gel phase transitions \cite{klauda2010update}. The area per head groups ($a_h$) of membranes from our simulations at all phases match well with the available literature \cite{KlaudaBBA:2014,KlaudaBBA:2018, KatsarasBBA:2011}(section II and table S2 of the SI). Furthermore, the governing mechanism of dynamical heterogeneity of supercooled water is unaffected by the choice of water model \cite{KimScienceAdvances:2017, galamba2016hydrogen}. \\
 Snapshots of the equilibrated bilayers at three phases are shown in figure \ref{snapshot} (a-c) and the bilayers at the remaining temperatures are shown in figure S1 of the SI (considered for the calculations of structural relaxations).  The $a_h$ drops abruptly from $308$ K to $292$ K and from $284$ K to $268$ K demonstrating the fluid (L$_{\alpha}$) to the ripple (P$_{\beta}$) to the gel ($L_{\alpha}$) phase transition temperatures (figure S2 of the SI). The ripple phase is identified from the existence of periodic interdigitated and non-interdigitated regions of lipid tails \cite{2005molecular}. The homogeneous thickness and interdigitation of the fluid phase ($324$ K-$303$ K) (figure \ref{snapshot} (d) and (g)) become periodically heterogeneous for the ripple phase ($292$ K-$284$ K) ((e) and (h)) which disappears for the gel phase ($273$ K-$218$ K). The region with smaller thickness is correlated to the higher interdigitation at all phases. The thickness and interdigitation of the remaining temperatures are shown in figure S4 in the SI (interdigitation). At the ripple phase, there is a coexistence of highly interdigitated ordered domain followed by a fully expanded disordered domain (shown in S4 in the SI). The ripple wavelength of $\sim4.2$ nm from figure \ref{snapshot} (e and h) matches well with that obtained from the spectral intensity of thickness and interdigitation fluctuations discussed in section II of the SI.

\begin{figure*}
\centering
  \includegraphics[width=0.9\linewidth]{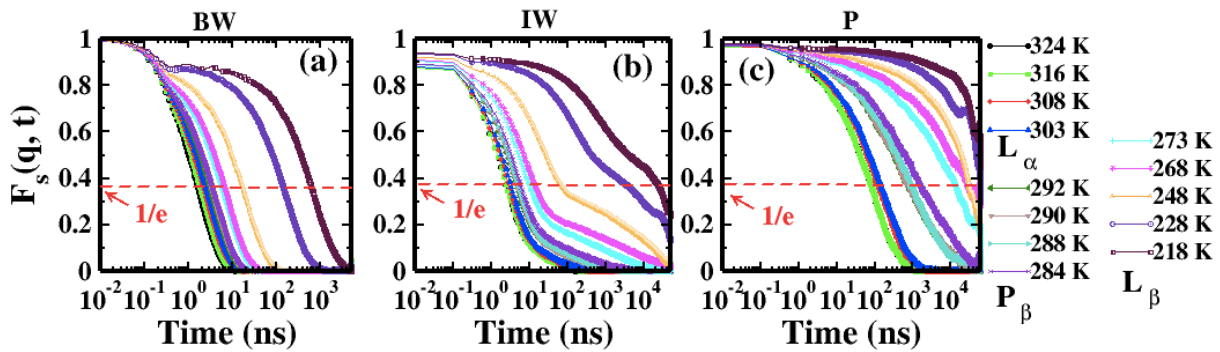}
  \caption{Self intermediate scattering function ($F_s(q,t)$) of a) BW, b) IW and c) P heads respectively at $\lambda$ = $0.5 nm$}
  \label{dynamics_BW_IW_P}
 \end{figure*}
\begin{figure*}
  \includegraphics[width=\linewidth]{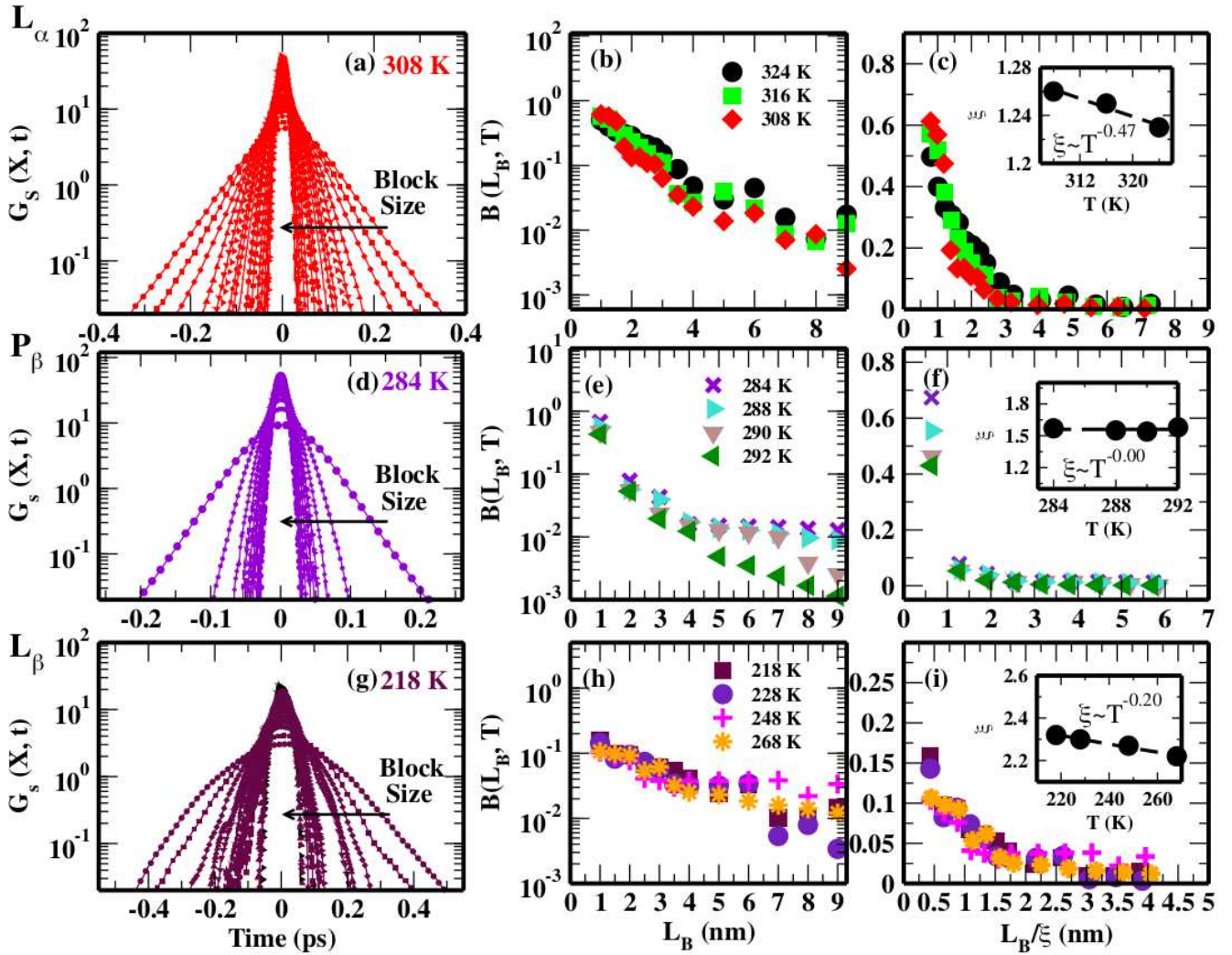}
  \caption{van Hove function, $G_s(X,t)$, of the IW for block lengths of $9-1$ nm for a) 324 K, d) 284 K and g) 218 K at the fluid, ripple and gel phases respectively. Binder cumulant of the blocked van Hove function at previously mentioned three temperatures in b), e) and h) respectively. Finite-size scaling of the binder cumulant by data collapse at three temperatures c), f), and i). Respective insets show the temperature dependence of the heterogeneity length scale, $\xi$, obtained from the data collapse of the binder cumulant at three phases.}
  \label{BC_different_phase}
 \end{figure*}

The homogeneous thickness and interdigitation of the fluid phase ($324$ K-$303$ K) (figure \ref{snapshot} (d) and (g)) become periodically heterogeneous for the ripple phase ($292$ K-$284$ K) ((e) and (h)) which disappears for the gel phase ($273$ K-$218$ K). The region with smaller thickness is correlated to the higher interdigitation at all phases. The thickness and interdigitation of the remaining temperatures are shown in figure S4 in the SI (interdigitation). At the ripple phase, there is a coexistence of highly interdigitated ordered domain followed by a fully expanded disordered domain (shown in S4 in the SI). The ripple wavelength of $\sim4.2$ nm from figure \ref{snapshot} (e and h) matches well with that obtained from the spectral intensity of thickness and interdigitation fluctuations discussed in section II of the SI.
As the largest survival time of the water molecule near membranes at the highest temperature is found to be $\sim 100$ ps \cite{DebnathCPL:2022}, it is considered as the minimum residence time of the IW at all temperatures. Thus, if a water molecule is present within $\pm 0.3$ nm of the lipid P head atoms {\it continuously} for $100$ ps, it is classified as an IW (section III of the SI). To find out the structural relaxation, self intermediate scattering functions (SISF) of the BW, IW and lipid heads are calculated (section IV of the SI). The initial decay of the SISF (figure \ref{dynamics_BW_IW_P} a)-c)) is due to the ballistic or cage motion followed by the emergence of Boson peaks at lower temperatures ($228 - 218$ K), as observed commonly in supercooled liquids. The time at which $F_s(q,t)=1/e$, is referred to as the $\alpha$ relaxation time-scale, $\tau_\alpha$. With lowering the temperature, $\tau_\alpha$ start growing for the BW, the IW and the P heads. The growths are significant for the IW and the P once the temperature goes below $273$ K and the membrane is at the gel phase. The growing $\tau_\alpha$ of the IW and the P heads are strongly correlated at all three phases of the membranes even at super-cooled temperatures (figure S5 in the SI).
 Block analysis has been employed on the membrane interface with $18\times18$ nm$^2$ surface area (Table S1 of the SI). The non-Gaussian van Hove distributions, $G_s(X,t)$, of the IW on the membrane surface become more and more Gaussian as the $L_B$ increases (see figure \ref{BC_different_phase} a), d), and g) for the fluid, ripple, and gel phases, respectively and figure S6 of the SI for remaining temperatures). Binder cumulant, $B (L{_B}, T)=0$ for higher temperatures and larger blocks as the underlying distribution is Gaussian. With decreasing block size, the values of the $B (L{_B}, T)$ increase at all temperatures due to larger deviations from Gaussianity seen in figure \ref{BC_different_phase} b), e), and h) for three phases of the membrane. As the binder cumulant is a scaling function of the underlying correlation length, it is assumed to follow the relation of $B (L{_B}, T)= f[L_{B}/\xi (T)]$. Thus, $B (L{_B}, T)$ at different temperatures are subjected to a data collapse and plotted in figure \ref{BC_different_phase} c), f) and i) for the fluid, ripple, and gel phases, respectively. The insets of figure \ref{BC_different_phase} depict the temperature dependence of the scaling length scale, $\xi$, known as heterogeneity length scale. For the fluid L$_\alpha$ phase, $\xi$ increases with a decrease in $T$. The dependence disappears for the ripple P$_\beta$ phase unlike the glassy liquids.\\

   \begin{figure}
  \centering
  \includegraphics[width=\linewidth]{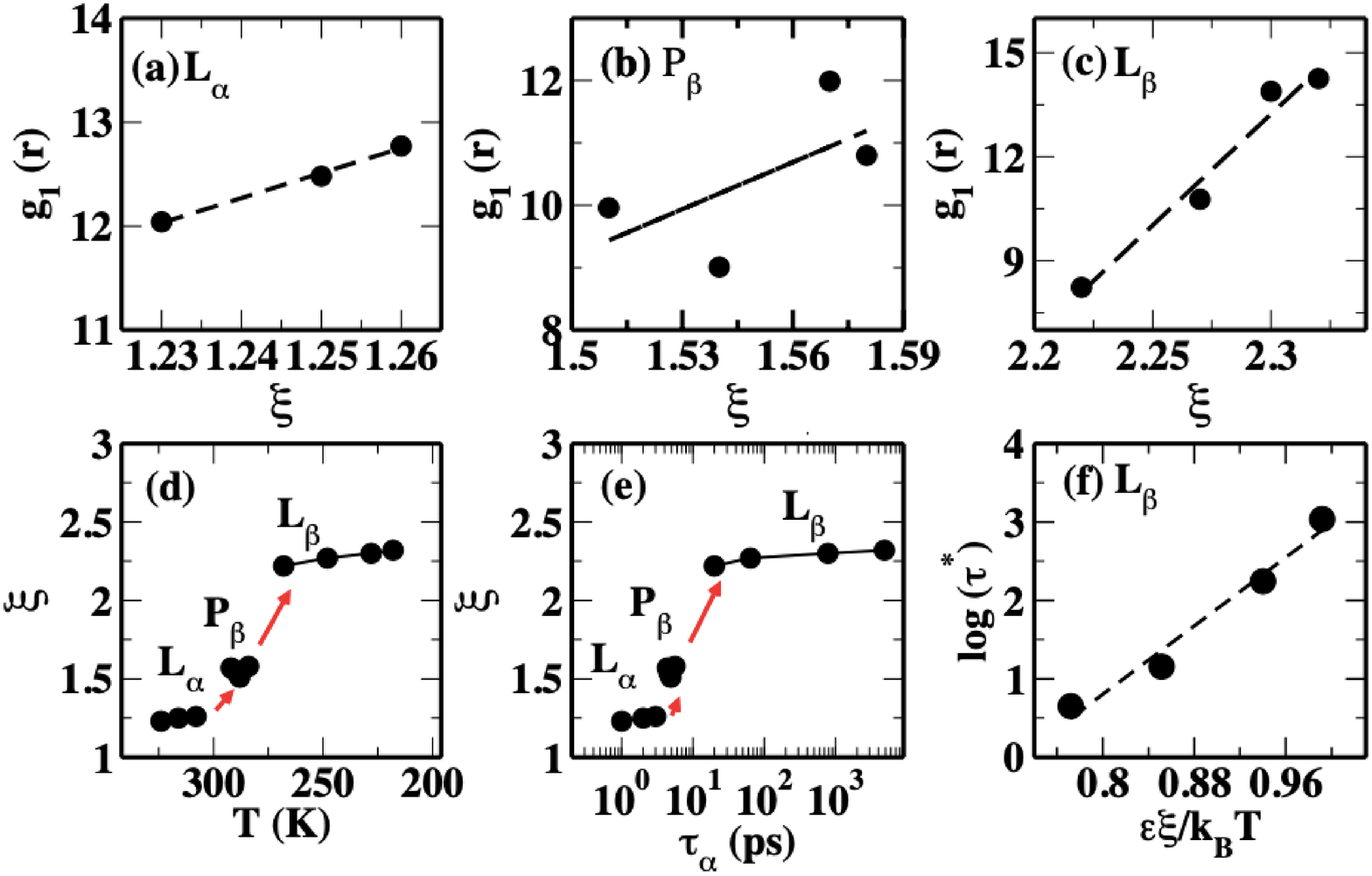}
  \caption{$g_1$ versus $\xi$ for a) L$_\alpha$, b) P$_\beta$, and c) L$_\beta$ phase. $g_1$ follows power law with $\xi$ where the exponents are $2.41$, $3.66$, and $13.24$ for three respective phases. d) Temperature dependence of $\xi$ of the IW showing sudden growth in $\xi$ at the phase transition temperatures. e) Dependence of log $\tau^*$ on $\epsilon\xi/k_B T$ follows an activated dynamic scaling for the L$_\beta$ phase. Here, $\tau^* = \tau_{\alpha} \sigma^{-1} \sqrt{\epsilon/M}$. $\sigma, \epsilon$, and $M$ of oxygens of water molecules are considered for the calculation.}
  \label{correlation}
 \end{figure}

 The values of $\xi$ at the ripple phase are close to the half of the ripple wavelengths as obtained from the spectra of thickness or interdigitation at all temperatures (figure S7 of the SI). This signifies that the periodic patch or domain of the ripple of the membrane can be captured by the heterogeneity length scale of the IW which do not vary with temperature within the same phase. The $T$ dependency of the $\xi$ is very weak at the gel phase. The heterogeneity length scale follows a power law behavior with the amplitude of the first peak of the radial distribution functions, $g_1$ \cite{DebnathCPL:2022} of the IW at each of the three phases of membranes (figure \ref{correlation} a), b) and c)) suggesting that the correlation length scale is structure dominated. Figure \ref{correlation} d) shows that the change in $\xi$ is minor within a given phase but the length scale significantly grows once there is a phase transition.
However, within a given phase, there is a drastic slow-down in the $\tau_\alpha$ with decreasing temperature without a significant change in $\xi$ (figure \ref{correlation} e). On the contrary, there are sharp changes in $\xi$, across the membrane phase transitions. Thus the heterogeneity time scales are slowed down mainly by supercooling which is not accompanied by a similar growth in the heterogeneity length scales. The large growth in the length scales is governed by the membrane phase transitions. The relaxation time scale, $\tau_{\alpha}$ of the IW obeys the relation, $\tau_{\alpha}\sim e^{a\xi/T},~a=$constant, solely near the gel phase (figure \ref{correlation} f)) which are not valid at the fluid or the ripple phase (figure S8 of the SI). \\

In conclusion, the study uses $11.55 \mu$s all-atom molecular dynamics simulations to correlate and quantify the spatio-temporal heterogeneities of the IW to the DMPC lipid membranes upon fluid-to-ripple-to-gel phase transitions.
Although there is a $\alpha$-relaxation time lag between the IW and the lipids, the lipid heads and the IW are strongly correlated to each other at all three phases with a strong enhancement in dynamical heterogeneity upon supercooling. The length scales of spatially heterogeneous dynamics of the IW are quantified using a block analysis technique, and then correlated to the lateral organizations of the membranes across phase transitions. One-dimensional van Hove correlation functions of the IW shows a shift from the non-Gaussianity to the Gaussianity once the block size increases at each phase of the membrane. The deviations from the Gaussianity are estimated by Binder cumulant which increases with lower block size at all temperatures.
Remarkably, the size of the rippling of the membrane can be obtained from the heterogeneity length scale of the IW. The length scale does not change significantly upon supercooling at the gel phase and follows an activated dynamical scaling with the relaxation time scale. This observation is in similar line to what is predicted by the random-field Ising model \cite{Young_98}. The drastic growth in the $\xi$ is dominated by the membrane phase transitions and the drastic slow-down in the relaxation time scale is dictated by the supercooling. Thus, the dynamic cross-over in the heterogeneity length scale of the IW will provide a new avenue to identify the micro-domain which can be critical for membrane functionalities. Since probing nano-domains in biological membranes are extremely challenging and no current biophysical tool is available yet to this end \cite{jacobsonNature:2007}, our method opens up a new direction to characterize functionally relevant nano-domains of membranes through the heterogeneity length scale of the IW. As per our knowledge, for the first time, a systematic analysis is performed to estimate the {\it spatio-temporal} correlations of the hydration water to the model lipid membranes across phases relevant to physiological and low-temperature stressed conditions. The findings can help us to better understand bio-protection processes under extreme conditions and will aid future research on domain-associated transport and signaling in biomembranes at low temperatures.\\
\\
See the supplementary material for the simulation methods including tables, and texts explaining the identification of the interface water, survival probability, bilayer properties, and figures. The supplementary files are available at\\
\\
A.D. acknowledges the financial support of the grant SERB CRG/2019/000106. S.M is thankful to Dr. Indrajit Tah for useful discussions and to Prof. Jeffery B. Klauda for providing the pdb of the equilibrated DMPC bilayer (with 72 lipids) at $273$ K.\\
\\
{\textbf{AUTHOR DECLARATIONS}}\\
{\bf{Conflict of Interest}}\\
The authors declare no competing interest.\\
{\bf{Author Contributions}}\\
ddS.M. performed the molecular dynamics simulations, analysis and wrote the original manuscript. S.K. supervised the project and manuscript editing. A.D. conceived the project, manuscript writing, project supervision, and  funding acquisition.\\
{\bf{DATA AVAILABILITY}}\\
The data that support the findings of this study are available within the article and its supplementary material.
%%%%%%%%%%%%%%%%%%

\bibliographystyle{unsrt}
\bibliography{paper}

%\end{thebibliography}
\end{document}